\documentclass[aps,showpacs,superscriptaddress,preprint]{revtex4}%
\usepackage{amsfonts}
\usepackage{amsmath}
\usepackage{amssymb}
\usepackage{graphicx}%
\begin{document}
\title{Magnetic quantum oscillations for the surface states of topological insulator
Bi$_{2}$Se$_{3}$}
\author{Zhigang Wang}
\affiliation{LCP, Institute of Applied Physics and Computational Mathematics, P.O. Box
8009, Beijing 100088, People's Republic of China}
\author{Zhen-Guo Fu}
\affiliation{State Key Laboratory for Superlattices and Microstructures, Institute of
Semiconductors, Chinese Academy of Sciences, P. O. Box 912, Beijing 100083,
People's Republic of China}
\affiliation{LCP, Institute of Applied Physics and Computational Mathematics, P.O. Box
8009, Beijing 100088, People's Republic of China}
\author{Shuang-Xi Wang}
\affiliation{State Key Laboratory for Superlattices and Microstructures, Institute of
Semiconductors, Chinese Academy of Sciences, P. O. Box 912, Beijing 100083,
People's Republic of China}
\affiliation{LCP, Institute of Applied Physics and Computational Mathematics, P.O. Box
8009, Beijing 100088, People's Republic of China}
\author{Ping Zhang}
\thanks{Corresponding author; zhang\_ping@iapcm.ac.cn}
\affiliation{LCP, Institute of Applied Physics and Computational Mathematics, P.O. Box
8009, Beijing 100088, People's Republic of China}
\affiliation{Center for Applied Physics and Technology, Peking University, Beijing 100871,
People's Republic of China}


\begin{abstract}
We study quantum oscillations of the magnetization in Bi$_{2}$Se$_{3}$(111)
surface system in the presence of a perpendicular magnetic field. The combined
spin-chiral Dirac cone and Landau quantization produce profound effects on the
magnetization properties that are fundamentally different from those in the
conventional semiconductor two-dimensional electron gas. In particular, we
show that the oscillating center in the magnetization chooses to pick up
positive or negative values depending on whether the zero-mode Landau level is
occupied or empty. An intuitive analysis of these new features is given and
the subsequent effects on the magnetic susceptibility and Hall conductance are
also discussed.

\end{abstract}
\maketitle

Magnetic oscillation, which was first predicted by Landau in 1930 \cite{r1},
has been a focus in the condensed matter physics filed. One important reason
is that the de Haas-van Alphen (dHvA) oscillations of magnetization provide a
vigorous technique to study the properties of carriers around the Fermi
surface. Especially in the last decade, thanks to the tremendous advances in
microscopic semiconductor technology, the challenge encountered in the
measurement of weak magnetization signal has been largely prominently
overcome, and the magnetic dHvA oscillations in the two-dimensional electron
gas (2DEG) systems have thus been extensively studied. For instance, Meinel
\textit{et al. }\cite{r2,r3,r4} developed dc superconducting quantum
interference device magnetometers to study the dHvA oscillations in
high-mobility semiconductor 2DEG. Schwarz \textit{et al.} \cite{r5,r6,r7}
studied the dHvA oscillations by using micromechanical cantilever
magnetometers. Besides the purely orbital part, prominently, the influence
\cite{Wang} of the weak Rashba spin-orbit interaction (SOI) on the dHvA
oscillations in the magnetization of the semiconductor 2DEG can also be
effectively determined in experiment \cite{Grundler}, which therefore opens a
new door to measurement of the spintronic parameters in semiconductor heterostructures.

\begin{figure}[ptb]
\begin{center}
\includegraphics[width=0.4\linewidth]{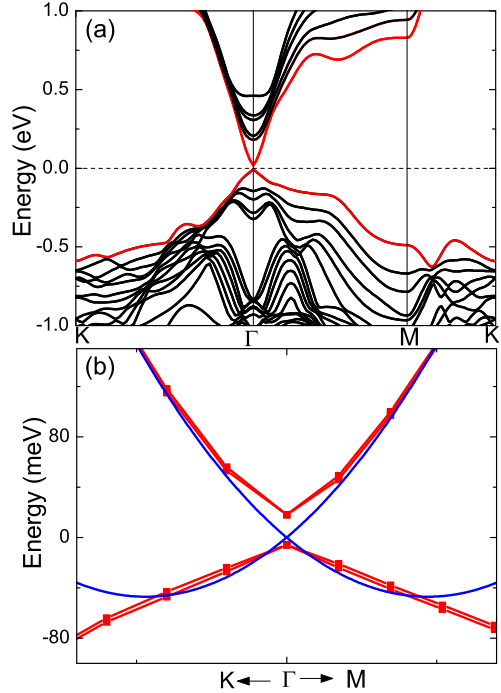}
\end{center}
\caption{(Color online) (a) The \textit{ab initio} calculated band structure
of the Bi$_{2}$Se$_{3}$(111) thin film with the thickness of six quintuple
layers. The red lines indicate the surface states while the black lines
correspond to the bulk bands. (b) The fitting curves (blue lines) around the
$\Gamma$ point with the model Hamiltonian [Eq. (1)]. }%
\label{Band}%
\end{figure}

In the above-mentioned conventional semiconductor 2DEG systems, in which the
electron motion is dominated by its orbital part, i.e., the magnetization
oscillation mainly comes from the response of the electron $k$-quadratic
kinetic energy to the external magnetic field. Although sometimes other
factors than the pure kinetic energy, such as the spin-orbit interaction
\cite{Wang}, have been taken into account, these factors in conventional
semiconductor 2DEG systems play only a minor role. For example, they can
result in a beating mode superposed onto the main dHvA oscillation pattern
\cite{Grundler}. This situation, however, will be greatly changed by very
recent theoretical finding \cite{Zhang1} and experimental verification
\cite{Zhang2} of the topological insulators (TIs) with strong spin-orbit
interaction. As a new state of matter as first addressed by Kane and Mele
\cite{Kane1}, the subject of time-reversal invariant TIs has attracted great
attention in condensed-matter physics. Several three-dimensional (3D) solids,
such as Bi$_{1-x}$Sb$_{x}$ alloys, Bi$_{2}$Se$_{3}$-family crystals, have been
identified \cite{Fu2007,Hsieh,Zhang2009,Xia2009,Chen2009} to be strong TIs
possessing anomalous band structures characterized by a \textbf{Z}$_{2}%
$-valued topological invariant \cite{Kane1,Kane2}. This invariant, called
$\nu_{0}$, counts the number of topologically protected gapless surface states
(modulo 2). A non-zero invariant means that the surface of 3D TIs will be
metallic. Instead of the conventional semiconductor 2DEG systems, the energy
scale for the surface states of these 3D TIs is dominated by the $k$-linear
spin-orbit interaction instead of the parabolic kinetic energy. As a result,
it is expected that the magnetic response properties of these topological
surface states are fundamentally different from those of the conventional 2DEG.

Inspired by this observation, as well as by the recent experimental
observation of the Landau quantization of the surface states of Bi$_{2}%
$Se$_{3}$ \cite{Xue2010}, in this paper we study the electron magnetic
oscillations of these surface states. Specially, we consider the surface
states of Bi$_{2}$Se$_{3}$. The first-principles surface band structure of
Bi$_{2}$Se$_{3}$ is calculated by \ a simple supercell approach with
spin-orbit coupling included and shown in Fig. \ref{Band}(a) along the
high-symmetry lines ($\mathbf{\Gamma}\mathtt{\rightarrow}\mathbf{M}$,
$\mathbf{M}\mathtt{\rightarrow}\mathbf{K}$, and $\mathbf{K}\mathtt{\rightarrow
}\mathbf{\Gamma}$) in the surface Brillouin zone. In obtaining Fig.
\ref{Band}(a), here we have used Vienna \textit{ab-initio} simulation package
(VASP) \cite{VASP}. The (Perdew-Burke-Ernzerhof) PBE \cite{PBE} generalized
gradient approximation and the projector-augmented wave potential \cite{PAW}
were employed to describe the exchange-correlation energy and the electron-ion
interaction, respectively. The SOI, which has been confirmed to play an
important role in the electronic structure of Bi$_{2}$Se$_{3}$, was included
during the calculation. The cutoff energy for the plane wave expansion was set
to 300 eV. During the calculation, the experimental lattice constants
\cite{constant} were adopted, i.e., $a$=$4.143$ \r{A}, $c$=$28.636$ \r{A},
with the internal parameter optimized automatically. The Bi$_{2}$Se$_{3}$(111)
surface was modeled by a slab composing of six quintuple layers (QLs) and a
vacuum region of 20 \r{A}. Integration over the Brillouin zone was done using
the Monkhorst-Pack scheme \cite{Monkhorst1976} with 10$\times$10$\times$1 grid
points for surface calculations. The structures of bulk and slab were fully
optimized until the maximum residual ionic force were below 0.01 and 0.02
eV/\r{A}, respectively. From Fig. \ref{Band}(a) two chiral surface states are
clearly seen to connect the conduction band and valence band, and cross each
other to form a single Dirac-type contact at the $\mathbf{\Gamma}$ point and
aligning with the Fermi energy. The intrinsic defects such as the
substitutional Bi defects at Se sites or the Se vacancies will play a role of
$n$-doping, and consequently shift the Fermi level above the Dirac point. Note
that due to the difference in the local symmetry between the top and bottom
surfaces, there can develop an observable asymmetric charge distribution on
the two surfaces if the sample is thin enough. This fact sometimes can open a
small gap between the two spin chiral surface states as shown in Fig.
\ref{Band}(b) which presents an enlarged version of the surface bands around
the Dirac cone. With increase of the thickness of the film, however, this
asymmetry-induced Dirac gap tends to vanish, and this actually corresponds to
the recent Landau quantization experiment, in which the used epitaxial
Bi$_{2}$Se$_{3}$ is as thick as $50$ QLs.

We report the calculated magnetization of the electrons on Bi$_{2}$Se$_{3}$
surface as a response to the external magnetic field. It is found that the
magnetization oscillations in the present system differs from the traditional
2DEG by the fact that the dHvA oscillating center in Bi$_{2}$Se$_{3}$ departs
from the well known (zero) value in the semiconductor 2DEG system. This
departure has an intimate relation with the different Landau spectrum
structures in these two kinds of systems. It is well known that the Landau
levels (LL's) in the traditional 2DEG obey the $B(n$+$1/2)$ rule with $B$
being the external magnetic field and $n$ the LL index. However, the energy
spectrum of the surface states in Bi$_{2}$Se$_{3}$ approximately obeys a
$\sqrt{nB}$ rule. It is this difference in LLs that distributes differently in
the two components of the magnetization, and eventually result in the
different magnetic properties in these two systems. Furthermore, we have shown
that the zero-mode LL plays a key role in determining the magnetization
behavior in the TI surface systems.

\begin{figure}[ptb]
\begin{center}
\includegraphics[width=0.4\linewidth]{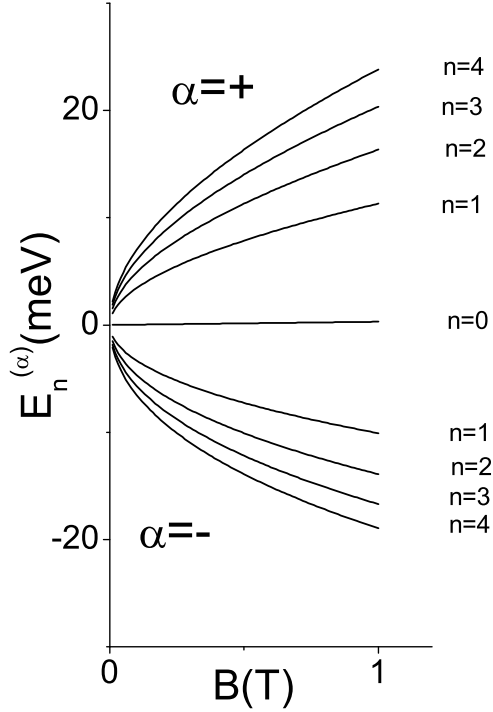}
\end{center}
\caption{The Landau levels in the Bi$_{2}$Se$_{3}$(111) surface system as
functions of the external magnetic field $B$.}%
\label{E1}%
\end{figure}

The Hamiltonian describing the gapless surface states of Bi$_{2}$Se$_{3}$ can
be approximately written as%
\begin{equation}
H(\mathbf{k})=\gamma k^{2}+\hslash v_{F}\left(  k_{x}\sigma_{y}-k_{y}%
\sigma_{x}\right)  , \label{Ham}%
\end{equation}
where $v_{F}$ is the Fermi velocity and $\mathbf{\sigma}$ are the Pauli
matrices for surface-state electron spins. Note that this Hamiltonian has the
same form as that of the conventional 2DEG system with Rashba spin-orbit
coupling \cite{Wang}. However, the intrinsic difference between these two
kinds of systems is that the $k$-linear spin-orbit interaction is primary to
the TI surface states, while the parabolic term is dominant in the
conventional 2DEG. Although it is very simple, the Hamiltonian (\ref{E1}) is a
general one, which can satisfactorily describe the gapless surface states of
Bi$_{2}$Se$_{3}$ near the Dirac point. This satisfaction is particularly
obvious for the upper part of the Dirac cone (electron part), as shown in Fig.
1(b), where Eq. (\ref{E1}) is used to fit the first-principles result, which
gives $\gamma\mathtt{=}2.1\mathtt{\times}10^{2}$ meV nm$^{2}$ and $\hslash
v_{F}\mathtt{=}200$\texttt{ }meV nm (namely, $v_{F}$=$3.04\times10^{5}$ m/s).
The effective mass $m^{\ast}$ is then obtained as $0.18m_{e}$ according to
$\gamma$=$\hslash^{2}/2m^{\ast}$, where $m_{e}$ is the mass of a free
electron. The lower surface states (hole part) is not well described by Eq.
(\ref{E1}) and a better fitting needs higher $k$-order corrections. For
simplicity, and for the reason that we only concern the $n$-doping, here we
neglect $\mathcal{O}$($k^{3}$) terms.

Let us now consider an external magnetic field $\mathbf{B}$=$B\hat{z}$ being
perpendicular to the surface. Taking the Landau gauge for the vector
potential, $A_{x}$=$By$ and $A_{y}$=$0$, and the transform $\hslash
\mathbf{k}\mathtt{\rightarrow}\mathbf{\Pi}$=$\hslash\mathbf{k}$+$e\mathbf{A}$,
one can obtain the following Hamiltonian
\begin{equation}
H=\frac{\mathbf{\Pi}^{2}}{2m^{\ast}}+v_{F}\left(  \Pi_{x}\sigma_{y}-\Pi
_{y}\sigma_{x}\right)  -\frac{1}{2}g_{s}\mu_{B}B\sigma_{z}, \label{H2}%
\end{equation}
where $g_{s}$ is the effective magnetic factor of the surface electron and
$\mu_{B}$ is the Bohr magneton. For Bi$_{2}$Se$_{3}$-family (111) surfaces,
the value of $g_{s}$ is approximately $8.0$ \cite{Liu2010}. Taking the fact
that the system is translation invariant along the $x$ axis and therefore the
wave number $k_{x}$ along this direction is a good quantum number, the
Hamiltonian (\ref{H2}) can be rewritten as%
\[
H=\hslash\omega_{c}\left(  a^{\dag}a+\frac{1-g\sigma_{z}}{2}+i\sqrt{2}%
\eta(a\sigma_{-}-a^{\dag}\sigma_{+})\right)  ,
\]
where $\sigma_{\pm}$=$(\sigma_{x}\pm i\sigma_{y})/2$, $\omega_{c}$%
=$eB/m^{\ast}$, $\eta$=$v_{F}m^{\ast}l_{B}/\hslash$ is the effective Rashba
spin-orbit coupling with $l_{B}$=$\sqrt{\hslash/eB}$ being the magnetic
length, $g$=$g_{s}m^{\ast}/2m_{e}$, and $a$=$[y$+$(\hslash k_{x}%
+ip_{y})/eB]/\sqrt{2}l_{B}$ is the usual harmonic oscillator operator. The LLs
are then given by
\begin{equation}
E_{n}^{(\pm)}=\hslash\omega_{c}\left(  n\pm\frac{1}{2}\sqrt{(1-g)^{2}%
+8n\eta^{2}}\right)  \label{d}%
\end{equation}
with $n$=$1$, $2$, $\cdots$. The $n$=$0$ LL only has the \textquotedblleft%
$+$\textquotedblright\ branch, $E_{0}^{(+)}$=$\frac{|1-g|}{2}\hslash\omega
_{c}$. The corresponding two-component eigenstates for $E_{n}^{(\pm)}$ are
given by
\begin{equation}
|n\rangle^{(\pm)}=\left(
\begin{array}
[c]{c}%
\cos\theta_{n}^{(\pm)}|n\rangle\\
i\sin\theta_{n}^{(\pm)}|n-1\rangle
\end{array}
\right)  ,
\end{equation}
where $|n\rangle$ is the eigenstate of the $n$th LL of a free two-dimensional
spinless electron. Here, $\tan\theta_{n}^{(\pm)}$=$-u_{n}\pm\sqrt{1+u_{n}^{2}%
}$ with $u_{n}$=$(1-g)/\sqrt{8n}\eta$ when $n\mathtt{>}0$ and $\theta
_{0}^{(+)}$=$0$ for $n$=$0$. It is interesting to see that the $n$=$0$ LL has
the fully polarized spin along the $z$ direction. Figure \ref{E1} plots the
LLs as functions of the magnetic field. Note that although the LL equation
(\ref{d}) for Bi$_{2}$Se$_{3}$ surface states has a similar form with the
conventional spin-orbit coupled semiconductor 2DEG \cite{Wang}, these two
systems are fundamentally different by the amplitudes of the physical
parameters. For the former the dimensionless parameter $\eta\mathtt{\gg}1$
while for the latter $\eta\mathtt{\ll}1$. Actually, for Bi$_{2}$Se$_{3}$, the
Se-terminated (111) surface lattice constant is $a$=$4.143$ \AA . With this
knowledge and through a normal fitting process, we obtain that at the external
magnetic field $B$=$1$ T, $\hslash\omega_{c}$=$0.61$ meV and $\eta$=$12.4$.
However, for a conventional 2DEG system with Rashba coupling, the
dimensionless parameter $\eta$ is typically in the range $0\mathtt{\sim}0.2$.
Based on this fact, the energy spectrum (\ref{d}) can be well approximated by
the dispersion relation
\begin{align}
E_{n}^{(\pm)}  &  =\pm\hslash\omega_{c}\sqrt{2n\eta^{2}+\frac{g^{2}}{4}}=\pm
v_{F}\sqrt{2ne\hslash B+\delta^{2}}\text{ \ (}n\neq0\text{)},\label{d1}\\
E_{0}^{(+)}  &  =\text{sgn}(g_{s})v_{F}|\delta|,\nonumber
\end{align}
where $\delta$=$g_{s}\mu_{B}B/2v_{F}$. Because the Zeeman splitting is much
smaller than the LL separations (for example, $g$=$0.72$ when $g_{s}$ takes
the value $8$, resulting in $\frac{1}{2}g_{s}\mu_{B}B$=0.13 meV at B=1 T),
thus the effect of the Zeeman term on the $n\mathtt{\neq}0$ LLs is very tiny
and can be safely neglected in considering the electron occupation of
$n\mathtt{\neq}0$ LLs. It is not so, however, for the $n$=$0$ LL. In fact, in
the absence of the Zeeman splitting, the Dirac-Landau energy spectrum
(\ref{d1}) is massless with a whole electron-hole symmetry and only half of
the zero mode is occupied by electrons in the case of $n$-doping. If the
Zeeman splitting is finite, the spectrum (\ref{d1}) is massive and the $n$=$0$
LL shifts upward or downward, depending on the orientation of the exchange
field (the sign of $g_{s}$). Correspondingly, this \textquotedblleft
zero\textquotedblright\ mode will be saturated by electrons for $g_{s}%
\mathtt{>}0$ or empty for $g_{s}\mathtt{<}0$, which, as shown in the following
discussion, will greatly influence the behavior of the magnetic response of
the system.

Now with the LL spectrum (\ref{d1}) [or the $k^{2}$-corrected LL spectrum
(\ref{d})], we study the magnetization of the surface electrons of Bi$_{2}%
$Se$_{3}$. The magnetization density is the derivative of the Helmholtz free
energy density with respect to $B$ at fixed electron density $\mathcal{N}$ and
temperature $T$, $M$=$-(\partial F/\partial B)|_{\mathcal{N},T}$. The
Helmholtz free energy density is given by%
\begin{align}
F(B,T)  &  =\mu\mathcal{N}-\frac{N_{\nu}}{\beta}\sum_{n=1}^{\infty}\ln\left[
1+e^{\beta(\mu-E_{n}^{(+)})}\right] \label{f}\\
&  -\frac{N_{0}}{\beta}\ln\left[  1+e^{\beta(\mu-E_{0}^{(+)})}\right]
,\nonumber
\end{align}
where $\beta$=$1/k_{B}T$, $N_{\nu}$=$1/2\pi l_{B}^{2}$ is the degeneracy for a
non-zero LL (namely, the number of electrons per unit area on a LL), and $\mu$
is the electron chemical potential. The second line in Eq. (\ref{f}) denotes
the contribution from the $n\mathtt{=}0$ LL and there exist three
possibilities for its contribution: (i) If this level is exactly a zero mode,
$E_{0}^{(+)}\mathtt{=}0$, then the system has the electron-hole symmetry and
half of the particles in the zero mode are electrons. In this case
$N_{0}\mathtt{=}N_{\nu}/2$; (ii) If the Zeeman splitting cannot be neglected
and the $g$ factor is positive as Eq. (\ref{d}) shows, then the $n\mathtt{=}0$
LL shifts upward and is fully accessible to electrons. In this case
$N_{0}\mathtt{=}N_{\nu}$; (iii) Otherwise, if the $g$ factor is negative, then
the $n\mathtt{=}0$ LL shifts downward and is unavailable to electron
occupation. In this case $N_{0}$=$0$. The $B$-dependent chemical potential
$\mu$ in Eq. (\ref{f}) is connected to the experimentally accessible electron
density $\mathcal{N}$, which is given by%
\begin{equation}
\mathcal{N}=N_{\nu}\sum_{n=1}^{\infty}f_{n}^{(+)}+N_{0}f_{0}^{(+)}%
\end{equation}
with $f_{n}^{(+)}$=$1/\left[  e^{\beta\left(  E_{n}^{(+)}-\mu\right)
}+1\right]  $ being the Fermi-Dirac distribution of the LL $E_{n}^{(+)}$. From
Eq. (\ref{f}) the electron magnetization density becomes%
\begin{align}
M  &  =-\left(  \sum_{n=1}^{\infty}N_{\nu}f_{n}^{(+)}\frac{\partial
E_{n}^{(+)}}{\partial B}+N_{0}f_{0}^{(+)}\frac{\partial E_{0}^{(+)}}{\partial
B}\right) \nonumber\\
&  +\left(  \frac{e}{h}\sum_{n=1}^{\infty}\frac{1}{\beta}\ln\left[
1+e^{\beta(\mu-E_{n}^{(+)})}\right]  +\frac{1}{\beta}\frac{\partial N_{0}%
}{\partial B}\ln\left[  1+e^{\beta(\mu-E_{0}^{(+)})}\right]  \right)
\nonumber\\
&  \equiv M^{(0)}+M^{(1)}. \label{m}%
\end{align}
The first part $M^{(0)}$ is the conventional contribution from the $B$
dependence of the LLs and thus denotes a diamagnetic response. The second part
$M^{(1)}$ comes from the $B$ dependence of the level degeneracy factor
$N_{\nu}$, thus describing the effect of the variation in the density of
states upon the magnetic field and denoting a paramagnetic contribution to the
total magnetization. Obviously, $M^{(0)}$ is negative while $M^{(1)}$ is
positive, the net result is an oscillation of the total magnetization $M$ as a
function of $B$.

\begin{figure}[ptb]
\begin{center}
\includegraphics[width=0.5\linewidth]{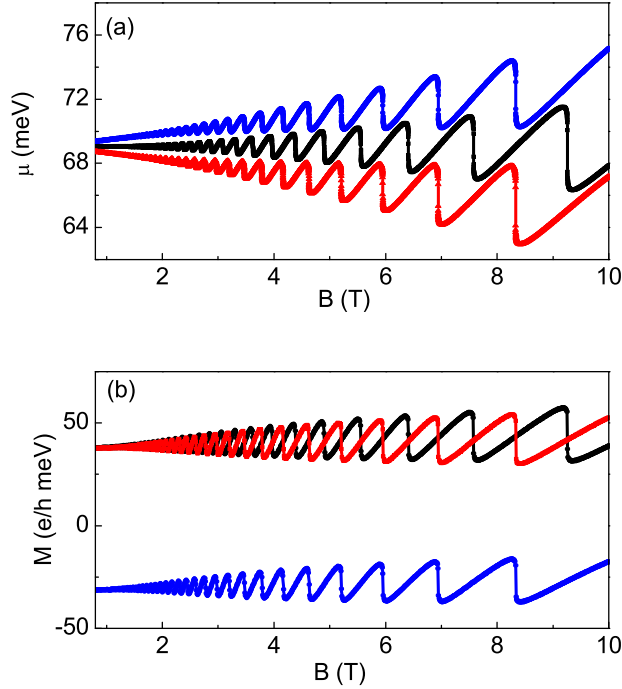}
\end{center}
\caption{(Color online) Magnetic field dependence of (a) chemical potential
$\mu$ and (b) magnetization $M$ in $n$-doped Bi$_{2}$Se$_{3}$(111) surface
system with electron density $\mathcal{N}$=$1.0\times10^{16}$ m$^{-2}$. The
temperature is set as $k_{B}T$=$0.3$ meV. The black, red, and blue curves
correspond to the cases that the zero mode is half filled, saturated, and
empty, respectively. }%
\label{f2}%
\end{figure}

We plot in Figs. \ref{f2}(a) and \ref{f2}(b) the magnetic dHvA oscillations of
the chemical potential and magnetization in Bi$_{2}$Se$_{3}$ for the
above-mentioned three cases of zero-mode filling. Comparing to the well-known
dHvA oscillating pattern in the conventional 2DEG, one can immediately obtain
three prominent features in the present TI surface system: (i) Although the
oscillating center of the chemical potential $\mu$ keeps a fixed value
unchanged by changing the external magnetic field $B$ when the $n$=$0$ LL is
exactly a zero mode, it linearly increases/decreases with $B$ when the $g$
factor is positve/negative; (ii) The oscillating center of the magnetization
$M$ keeps a \emph{non-zero} value unchanged by varying the external magnetic
field strength. This is totally different from those in the semiconductor
2DEG. It is well known that in the clean semiconductor 2DEG sample, the
oscillating center of the chemical potential $\mu$ keeps a constant value
unchanged and that of the magnetization $M$ keeps zero unchanged when varying
the magnetic field $B$; (iii) The magnetization for the case of empty zero
mode is fundamentally distinguished from the cases of saturated and
half-filling zero mode by a total sign inversion. In addition, the magnetic
oscillation patterns for the cases of saturated and half-filling zero mode are
out phase. Thus, Fig. \ref{f2} clearly reveals the fundamental role the zero
mode played in determining the magnetic response properties of the TI surface states.

For further illustration and to see the origin of the sign inversion in the
magnetization when the zero is changed from filling to unfilling, let us first
consider the case of saturated zero mode. In this case the $n$=$0$ LL is
occupied by electrons with the degeneracy $N_{0}\mathtt{=}N_{\nu}$. The
discrepancy in the dHvA oscillating patterns between the TI surface and the
semiconductor 2DEG comes from their different energy dispersion relations. The
former versus the external magnetic field obeys square root rule while the
latter obeys linear rule. It is well known that at zero temperature, the two
components of the magnetization in the conventional 2DEG turn to be $M^{(0)}%
$=$-\frac{e}{h}\sum_{n=0}^{\text{occu.}}E_{n}$ and $M^{(1)}$=$\frac{e}{h}%
\sum_{n=0}^{\text{occu.}}(\mu_{0}-E_{n})$, respectively. Here $\mu_{0}$ is the
zero-temperature Fermi energy and the LL $E_{n}^{(+)}\mathtt{\propto}B$. The
negative $M^{(0)}$ and the positive $M^{(1)}$ gives that the net result is an
oscillation of the total magnetization $M$ as a function of $B$. The
oscillation amplitude increases with $B$ and the oscillation center is zero.
However, because the LL $E_{n}\mathtt{\propto}\sqrt{B}$ for the present
system, at zero temperature the first component of $M$ turns to be $M^{(0)}%
$=$-\frac{e}{h}\sum_{n=0}^{\text{occu.}}\frac{E_{n}^{(+)}}{2}$, while the
second component is $M^{(1)}$=$\frac{e}{h}\sum_{n=0}^{\text{occu.}}(\mu
_{0}-E_{n}^{(+)})$. By comparison with those in the semiconductor 2DEG, one
can find that in the present TI surface system the diamagnetic contribution
($M^{(0)}$) is reduced. As a result, the oscillating center of the
magnetization is now a positive value for the case of saturated zero mode.
This simple comparison is not strict in mathematics, however, it affords an
intuitive explanation on the difference of the magnetization between the TI
surface and the semiconductor 2DEG.

In the case of empty zero mode, the $n$=$0$ LL is excluded and the first
component $M^{(0)}$ now becomes $M^{(0)}$=$-\frac{e}{h}\sum_{n=1}%
^{\text{occu.}}\frac{E_{n}^{(+)}}{2}$, while the second component becomes
$M^{(1)}$=$\frac{e}{h}\sum_{n=1}^{\text{occu.}}(\mu_{0}-E_{n}^{(+)})$.
Compared to the case of saturated zero mode, and considering $\mu
_{0}\mathtt{\gg}E_{0}^{(+)}$, one can find that the magnetization in the case
of the empty zero mode is smaller than the saturated case by a quantity
$\frac{e}{h}\mu_{0}$ and therefore becomes negative during its oscillations as
a function of $B$.

Note that the abrupt jump in the dHvA oscillation is on the high magnetic
field side of the sawtooth, which is special for our present choice of the
thermodynamic system. If the system is constrained to have constant chemical
potential, then the jump in the dHvA oscillation will move to the low magnetic
field side of the sawtooth, which has been confirmed by Meinel et al.
\cite{r2} in an experiment with the electron density $\mathcal{N}$ modulated
by applying a gate voltage to the sample.

\begin{figure}[ptb]
\begin{center}
\includegraphics[width=0.5\linewidth]{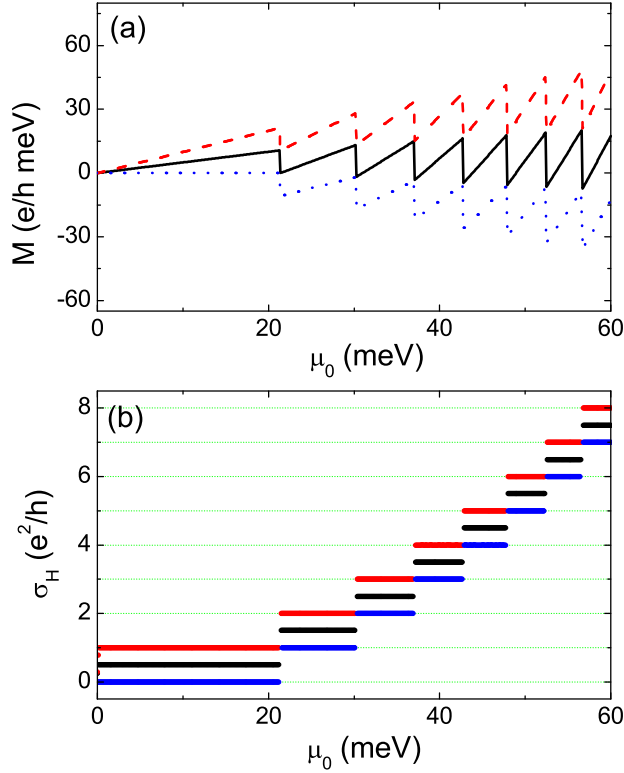}
\end{center}
\caption{(Color online) (a) Calculated magnetization as a function of the
Fermi energy $\mu_{0}$. The external magnetic field is chosen as $B$=$4$ T.
(b) The derived Hall conductance from $\partial M/\partial\mu_{0}$. The black,
red, and blue curves correspond to the cases that the zero mode is half
filled, saturated, and empty, respectively.}%
\label{fig2}%
\end{figure}

The above discussions on the dHvA oscillations of the magnetization focus on
the situation that the total number of electrons on the LL's is field
independent. Now we consider the magnetization properties in another situation
that the chemical potential is field independent. Figure \ref{fig2}(a) plots
the magnetization as a function of the zero-temperature chemical potential
(i.e., the Fermi energy) at $B$=$4$ T for three cases of zero-mode filling.
The dHvA oscillating patterns as a function of the chemical potential can be
observed from Fig. \ref{fig2}(a). There exist different kinds of patterns for
the dHvA oscillating center in different cases. When the $n$=$0$ LL is half
occupied by electrons, the dHvA oscillating center keeps zero unchanged by
increasing the chemical potential. When the zero mode is saturated/empty, the
dHvA oscillating center linearly increases/decreases by increasing the
chemical potential. A fact that should be addressed is that because the
chemical potential is tuned freely and independent with the external field,
there are no phase difference in the dHvA oscillations for different cases.
The corresponding $\partial M/\partial\mu_{0}$ are also calculated, from which
the Hall conductance $\sigma_{H}$ is obtained by combining the thermodynamic
relationship and Streda formula:%
\begin{equation}
\partial M/\partial\mu_{0}=\frac{1}{e}\sigma_{H}\text{.} \label{Ha}%
\end{equation}
The result of Hall conductance as a function of the Fermi energy is plotted in
Fig. \ref{fig2}(b), from which Hall plateaus can be clearly seen. The plateau
values of $\sigma_{H}$ depend on the zero-mode filling. If the $n$=$0$ LL is
half filled, the Hall conductance takes half-integer values of $\sigma_{H}%
$=($n$+$1/2$)$e^{2}/h$, as shown in Fig. \ref{fig2}(b) by black step lines. To
date, measuring the half-integer quantum Hall effect on the TI surfaces keeps
a challenging task, although the LLs have been recently observed
\cite{Xue2010}. If the zero mode is saturated, then $\sigma_{H}$=($n$%
+$1$)$e^{2}/h$, as shown in Fig. \ref{fig2}(b) by red step lines. Finally, if
the zero mode is empty, then $\sigma_{H}$=($n\mathtt{-}1$)$e^{2}/h$, as shown
in Fig. \ref{fig2}(b) by blue step lines. We note that the quantum Hall effect
in the TI surface system with finite sample size has also been discussed in
Ref. \cite{Shen}.

\begin{figure}[ptb]
\begin{center}
\includegraphics[width=0.6\linewidth]{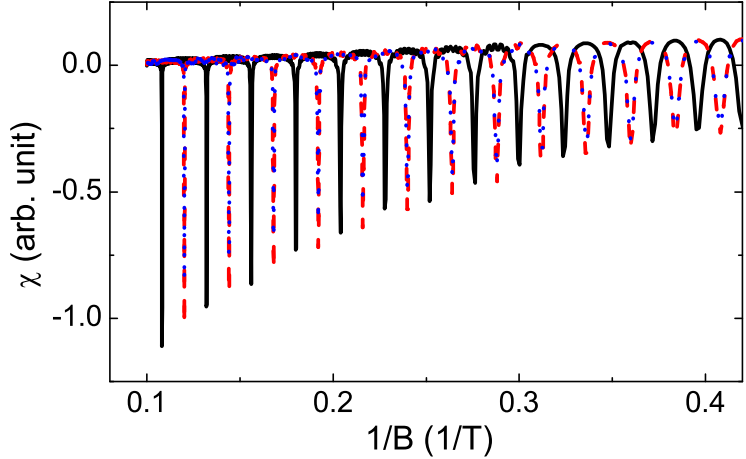}
\end{center}
\caption{(Color online) Calculated magnetic susceptibility $\chi$ as a
function of $1/B$. The parameters are the same as those in Fig. \ref{f2}. The
black, red, and blue curves correspond to the cases that the zero mode is half
filled, saturated, and empty, respectively.}%
\label{f42}%
\end{figure}

The information on the magnetic susceptibility $\chi$ of the TI surfaces,
which is defined as the derivative of the magnetization with respect to the
external magnetic field, $\chi$=$\partial M/\partial B$, can be easily
obtained with the knowledge of the magnetization [Eq. (\ref{m})]. The final
expression of the magnetic susceptibility is given by%
\begin{align}
\chi &  =\left[  \frac{e}{h}\sum_{n=1}^{\infty}f_{n}^{(+)}\left(
\frac{\partial\mu}{\partial B}-2\frac{\partial E_{n}^{(+)}}{\partial
B}\right)  +\frac{\partial N_{0}}{\partial B}f_{0}^{(+)}\left(  \frac
{\partial\mu}{\partial B}-2\frac{\partial E_{0}^{(+)}}{\partial B}\right)
\right] \\
&  -\left\{  \sum_{n=1}^{\infty}N_{\nu}\left[  \frac{\partial f_{n}^{(+)}%
}{\partial B}\frac{\partial E_{n}^{(+)}}{\partial B}+f_{n}^{(+)}\frac
{\partial^{2}E_{n}^{(+)}}{\partial B^{2}}\right]  +N_{0}\left[  \frac{\partial
f_{0}^{(+)}}{\partial B}\frac{\partial E_{0}^{(+)}}{\partial B}+f_{0}%
^{(+)}\frac{\partial^{2}E_{0}^{(+)}}{\partial B^{2}}\right]  \right\}
.\nonumber
\end{align}
Figure \ref{f42} plots the magnetic susceptibility in Bi$_{2}$Se$_{3}$ sample
as a function of the inverse magnetic field, $1/B$. Because the resonant (for
magnetic susceptibility) magnetic field only reflects the occupation of the
LLs, which is the same as that in the conventional 2DEG, it losses the message
on the oscillating center value of the magnetization. However, the difference
between the half-filled case and saturated/empty case of the zero mode is
still clearly revealed in this figure.

In summary, we have investigated the dHvA oscillations of the magnetization in
the Bi$_{2}$Se$_{3}$-family surface systems. Our results show that the dHvA
oscillating center of the magnetization maintains positive values when the
$n$=$0$ LL is fully occupied or half occupied. When this mode is empty, the
dHvA oscillating center changes a sign. These results are fundamentally
different from those in the conventional semiconductor 2DEG systems, in which
the dHvA oscillating center is at zero. We have given an intuitive analysis on
this difference, which turns to have an intimate relation with different forms
of the energy dispersions in these two kinds of systems. This can be reflected
by the fact that the LLs for the TI surfaces is proportional to $\sqrt{B}$
instead of $B$-linear dependence accommodated by the conventional 2DEG.
Furthermore, the essential role that the zero mode played has been illustrated
by the different behavior of the Hall conductance at three kinds of electron
occupation of this mode. We expect that the present results for the
topologically nontrivial features in the magnetic response of the TI surfaces
can be experimentally confirmed in the future topological magnetoelectric studies.

\textit{Note added}.--- While this work was completed, we were aware of an
experimental measurement \cite{Ando} of the magnetization for the topological
insulator Bi$_{1-x}$Sb$_{x}$ (0.07%
$<$%
$x$%
$<$%
0.22). Compared to Bi$_{2}$Se$_{3}$, the surface band structure of Bi$_{1-x}%
$Sb$_{x}$ alloy is much more complicated. Furthermore, in Bi$_{1-x}$Sb$_{x}$
the bulk band is often coupled with surface band during measurement. These
facts make it difficult to study the magnetic quantum oscillations that are
fully from the 2D surface states of Bi$_{1-x}$Sb$_{x}$ alloy. In spite of
these complicated facts, we expect that the phenomenon of large-amplitude dHvA
magnetic oscillations found in Ref. \cite{Ando} is closely related to our
theoretical finding in the present paper.

\begin{acknowledgments}
This work was supported by NSFC under Grants No. 90921003, No. 10904005 and
No. 60776063, and by the National Basic Research Program of China (973
Program) under Grant No. 2009CB929103.
\end{acknowledgments}

\end{document}